\documentclass[doublecol]{epl2} 
\usepackage{mathrsfs}
\usepackage[centertags]{amsmath}
\usepackage{amssymb}
\usepackage{amsthm}
\usepackage{amsmath}
\usepackage{CJK}
\usepackage{graphicx,subfigure,epstopdf}
\usepackage{indentfirst}
\usepackage{fancyhdr}
\usepackage{bm}
\usepackage{esint}
\usepackage{comment}
\newcommand{\ie}{\emph{i.e.}}
\newcommand{\eg}{\emph{e.g.}}
\newcommand{\ER}{Erd\H{o}s-R\'{e}nyi}

\title{Deriving an underlying mechanism for discontinuous percolation}

\author{W. Chen\inst{1,2,3},  Z. Zheng\inst{4}, \and R. M. D'Souza\inst{3,5}}
\shortauthor{W. Chen \etal}

\institute{                    
  \inst{1} School of Mathematical Sciences, Peking University - No.5 Yiheyuan Road, Haidian, Beijing 100871, China \\
  \inst{2} Institute of Computing Technology, Chinese Academy of Sciences - No.6 KeXueYuanNanLu, Haidian, Beijing 100190, China\\
  \inst{3} University of California, Davis CA, 95616 - One Shields Avenue, Davis, CA 95616, USA\\
  \inst{4} Beihang University - No.37 Xueyuan Road, Haidian, Beijing 100191, China.\\
  \inst{5} Santa Fe Institute - 1399 Hyde Park Road, Santa Fe, New Mexico 87501, USA
      }

\abstract{
Understanding what types of phenomena lead to discontinuous phase transitions in the connectivity of random networks is an outstanding challenge. 
Here we show that a simple stochastic model of graph evolution leads to a discontinuous percolation transition and we derive the underlying mechanism responsible: growth by overtaking.  Starting from a collection of $n$ isolated nodes, potential edges chosen uniformly at random from the complete graph are examined one at a time while a cap, $k$, on the maximum allowed component size is enforced.  Edges whose addition would exceed $k$ can be simply rejected provided the accepted fraction of edges never becomes smaller than a function which decreases with $k$ as 
$g(k) = 1/2 + (2k)^{-\beta}$. We show that if $\beta < 1$ it is always possible to reject a sampled edge
and the growth in the largest component is dominated by an overtaking mechanism leading to a  discontinuous transition.  If $\beta > 1$, 
once $k \ge n^{1/\beta}$, there are situations when a sampled edge must be accepted 
leading to direct growth 
dominated by stochastic fluctuations and a ``weakly" discontinuous transition. 
We also show that the distribution of component sizes and the evolution of component sizes are distinct from those previously observed and show no finite size effects for the range of $\beta$ studied. 
}

\begin{document}

\maketitle

\section{Introduction}
Percolation is a theoretical underpinning for analyzing properties of networks, including epidemic thresholds, vulnerability, and robustness~\cite{StaufferPercBook,NW1999,moorePRE,Cohen2000,Callaway2000,Newman2010}, with large-scale connectivity typically emerging in a smooth and continuous transition. 
A prototypical process 
begins from a collection of $n$ isolated nodes with edges connecting pairs of nodes sequentially chosen uniformly at random and added to the graph~\cite{ER}.  
A set of nodes connected by paths following edges is called a component, and the percolation phase transition corresponds to the first moment that there exists a component of size proportional to $n$ (\ie, a ``giant component"). 
A ``fixed choice" variant of the simple process has gained much attention in recent years~\cite{EPScience},
where instead of a single edge, at each discrete time step a fixed number of randomly selected candidate edges are examined together, but only the edge that maximizes or minimizes a pre-set criteria is added to the graph. The resulting percolation transition can be extremely abrupt, with a large 
discontinuous jump in connectivity observed in systems with sizes larger than any real-world network (\eg, tens of billions of nodes).  Yet in the $n \rightarrow \infty$ limit any  fixed choice graph evolution rule leads to a continuous transition~\cite{daCostaArxiv,RiordanWarnke,Grassberger,LeeKimPark} 
(which may actually be followed by a discontinuous jump arbitrarily close to the transition point~\cite{naglerPRX}).
Several models that lead to truly discontinuous percolation transitions are now known, {\it e.g.}~\cite{AraujoHerrmann,CD2011,restrictedER,AraujoPRL2011,GleesonPRL2011,ChoandKahngPRE2011,Schrenk2012,naturecommofZiff}, yet the underlying mechanisms 
are not fully 
understood. There are  many investigations underway to isolate essential ingredients that lead to a discontinuous transition such as cooperative phenomena~\cite{Bhizani}, hierarchical structures~\cite{naturecommofZiff},  correlated percolation~\cite{Schwarz}, and algorithms that explicitly suppress types of  growth~\cite{ChoKahngPRL11}. 

Here 
we show that a simple stochastic graph evolution 
process, that examines only one edge at a time, leads to a discontinuous transition and we analytically derive the simple underlying mechanism for this process: growth by overtaking.
The size of the largest component changes not by direct growth, but instead when two smaller components merge together and become the new largest. 
Overtaking
is a natural growth mechanism observed in a range of systems from industrial firms~\cite{overtakingBusiness} to ecosystems~\cite{overtakingEco,overtakingEcon}, 
where two smaller entities choose to cooperate (or merge) to gain competitive advantage over a previously larger entity. For the simple model studied here, we show that there is a control parameter (denoted $\beta$) that when small enough only allows significant growth by overtaking and leads to a discontinuous transition.  But once the parameter is large enough, substantial direct growth of the giant component is allowed leading to a continuous transition. 
We also show that the distribution of component sizes is distinct from any previously observed. Likewise, the time evolution of the components sizes is novel. Also novel are the lack of finite size effects 
 for the range of $\beta$ studied.



\section{Model}
The basic model we analyze was originally introduced by Bohman, Frieze and Wormald (BFW)~\cite{BF} and predates~\cite{EPScience}.
The BFW process is initialized with a collection of $n$ isolated nodes and a cap on the maximum allowed  component size set to $k=2$. Edges are then sampled one at a time, uniformly at random from the complete graph and either added to the graph or rejected following the algorithm in Table~\ref{BFWalg}. If an edge would lead to formation of a component of size less than or equal to $k$ it is accepted (and we move on to sample a new edge). Otherwise, check if  the fraction of accepted edges 
remains greater than or equal to a function $g(k)=1/2+(2k)^{-1/2}$. If so, 
the edge is simply rejected (and we move on to a new edge). If not, 
the cap is augmented to $k+1$ and we iterate the algorithm again. In other words, in this final case, $k$ is augmented by one repeatedly until either $k$ has increased sufficiently to accept the edge or $g(k)$ has decreased sufficiently that the edge can be rejected. 

\begin{table}[tb]
\caption{The BFW algorithm. At each step $u$, the selected edge $e_u$ is examined via this algorithm, where $u$ denotes the total number of edges sampled, $A$ the set of accepted edges (initially $A =\emptyset$), and $t=|A|$ the number of accepted edges. }
\begin{center}
\begin{tabular}{|l|}
\hline
{Set $l=$ maximum size component in $A \cup \{e_{u}\}$}\\

if $\left(l\leq k\right) \{$\\
 
 \ \ \ $A \leftarrow A \cup \{e_{u}\}$\\
 
 \ \ \ $u\leftarrow u+1.$ (Get next edge.)\}\\
  
   else if $\left(t  / u \ge g(k)\right) \{u \leftarrow u+1 $.  (Get next  edge.)\}\\
 
  else \{\ $k\leftarrow k+1$. Then repeat this block.\}\\ \hline

\end{tabular}
\label{BFWalg}
\end{center}
\end{table}%


\begin{figure}
\includegraphics[width=0.48\textwidth]{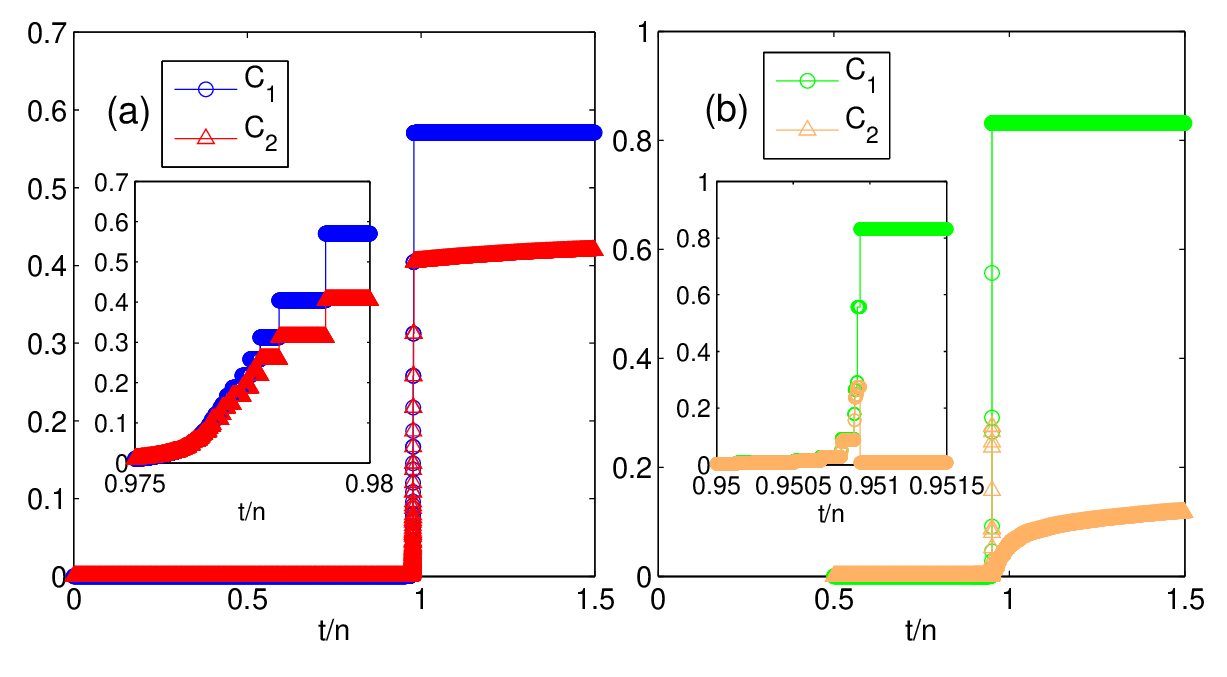}
\caption{
Evolution of $C_{1}$ and $C_{2}$ versus edge density, $t/n$, for $n=10^6$. (a) For $\beta=1/2$, two giant components emerge simultaneously. Inset  is the behavior in the critical region showing growth via the overtaking process when what was $C_1$ becomes $C_2$. 
(b) A typical realization for $\beta=2.0$. Inset shows direct growth, that  $C_{1}$ and $C_{2}$ merge together, and what was $C_3$ becomes the new $C_2$. 
}
 \label{fig:as}
 \end{figure}
 
\begin{figure*}
 
 \includegraphics[width=0.9\textwidth]{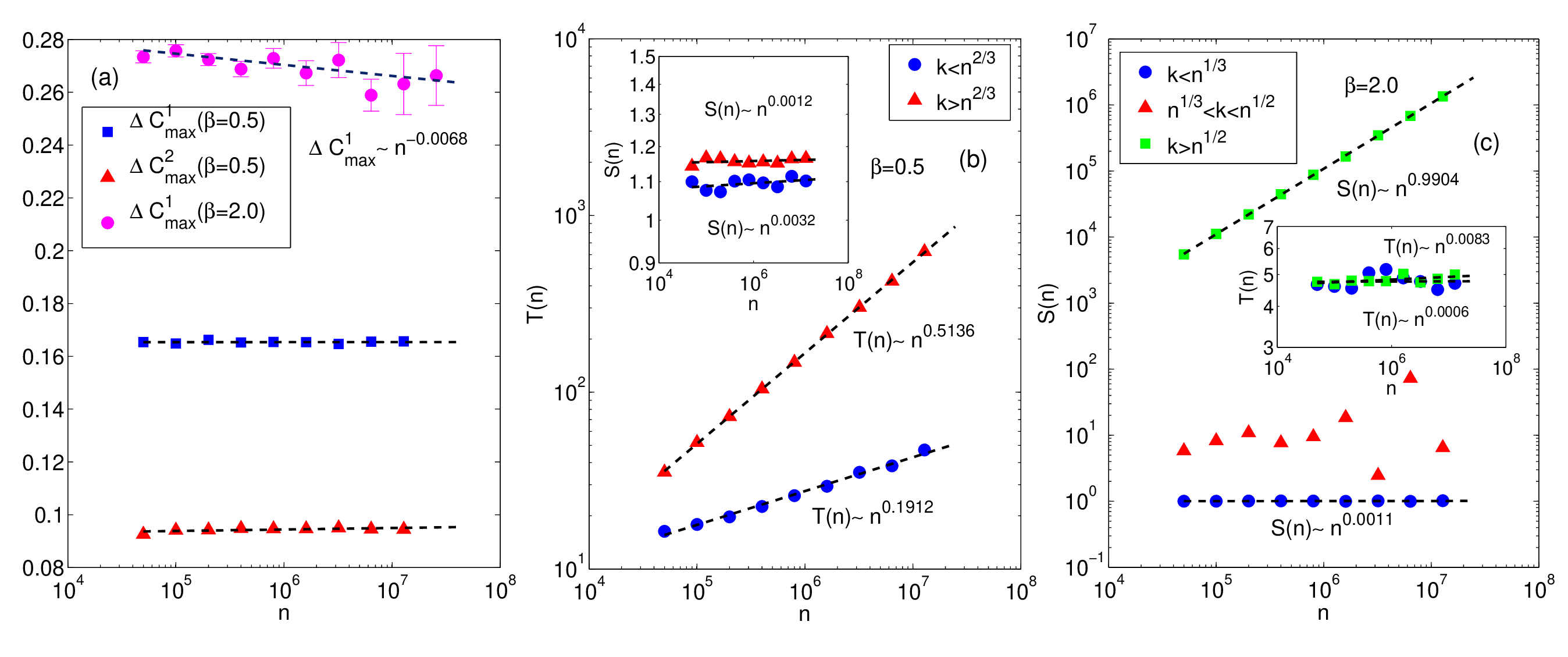}
 \centering
\caption{Slow convergence leads to a strongly discontinuous transition and growth by overtaking. (a) For $\beta=0.5$, $\Delta C^{1}_{\rm max}$ and $\Delta C^{2}_{\rm max}$ are independent of $n$ and both largest components emerge in strongly discontinuous phase transitions.  For $\beta=2.0$, $\Delta C^{1}_{\rm max}\approx 0.285 n^{-0.0068}$, showing a weakly discontinuous phase transition. 
 (b) For $\beta =0.5$, main plot is $T(n)$ the number of times $C_1$ undergoes {\tt direct growth} versus $n$, with the two regimes separated by $k = n^{1/(\beta+1)}$.   (Inset) $S(n)$, the average size of component that merges with $C_1$ during direct growth, is essentially a constant, $S(n)\approx 1.1$ (\ie, an isolated node), in both regimes. 
 (c) For $\beta =2.0$, main plot is $S(n)$, showing three regimes. For $k < n^{1/(\beta+1)}$ we observe $S(n)\approx 1$. The intermediate regime is noisy.  Then once $k > n^{1/\beta}$, random edges must be accepted at times and $S(n) \sim n^{0.9904}$ ($C_1$ merges with other essentially macroscopic components).  (Inset) $T(n)$ is essentially constant: $T(n) \approx 5$. 
All data points are the average over 30 to $100$ independent realizations (based on system size), with error bars smaller than the symbols unless otherwise indicated.}
 \label{fig:bs}
 \end{figure*}

Here we modify the original BFW function above such that $g(k)=1/2+({2k})^{-\beta}$, for $\beta \ge 0.5$, 
to analyze how the parameter $\beta$, which controls the rate of convergence of $g(k)$ to its asymptotic limiting value of $1/2$, affects the nature of the transition.  
Letting $C_i$ denote the fraction of nodes in the $i$th largest component, we show both analytically and via numerical investigation that for $\beta<1$ any significant growth in $C_1$ is dominated by an overtaking mechanism where smaller components merge together to become the new largest component, leading to 
a discontinuous transition.
In contrast, if $\beta>1$ significant direct growth of $C_1$ is allowed and the process is dominated by stochastic fluctuations, leading to a ``weakly" discontinuous transition that is likely continuous as $n \rightarrow \infty$. 
The typical evolution of $C_1$ and $C_2$ for $\beta=0.5$ and $\beta=2.0$ are shown in Figs.~\ref{fig:as}(a) and (b). 
(The simultaneous emergence of multiple stable giant components was shown in~\cite{CD2011}, but the underlying mechanism leading to the discontinuous transition, our current focus, was not identified.)

 
%
%
%
%
%
%
%

\section{Methods and Results}
We numerically measure $C_1, C_2$, and $C_3$ throughout the evolution for various $\beta \in[0.5,\infty]$, for a large ensemble of realizations and range of system sizes $n$. For each realization we define the critical point as the single edge $t_c$ whose addition causes the largest change in the value of $C_1$, with this largest change denoted by $\Delta C^1_{\rm max}$.  As shown in Fig.~\ref{fig:bs}(a), for $\beta=0.5$, $\Delta C^1_{\rm max}$ is independent of system size $n$ and discontinuous. The same holds for $\Delta C^2_{\rm max}$, the largest jump in $C_2$.
Yet, for $\beta=2.0$, $\Delta C^1_{\rm max}\approx 0.285 n^{-0.0068}$. With this scaling a system of size $10^{66}$ would have $\Delta C^1_{\rm max} \sim 0.1$. 
Following Ref.~\cite{Timme2010} we label this ``weakly" discontinuous, to describe the extremely slow decrease of jump size with $n$.

%
%

As discussed in~\cite{Timme2010}, whenever a single edge is added to the evolving graph, $C_1$ may increase due to one of three mechanisms:   (1) {\tt Direct growth}, when the largest component merges with a smaller one; (2) {\tt Doubling}, when two components both of fractional size $C_1$ merge (this is the largest increase possible); (3) {\tt Overtaking}, when two smaller components merge together to become the new largest. 
In~\cite{Timme2010} it is proven that if direct growth is strictly
prohibited up to the step when only two components remain in the system, then a strongly discontinuous transition ensues. 
We next show via analytic arguments that for our modified BFW process,  if $\beta < 1$ then throughout the subcritical regime direct growth only occurs when the largest component merges with an essentially isolated node and all significant growth is due to overtaking.  In contrast, if $\beta >1$ then the initial evolution is the same, but once $C_1 \sim n^{1/\beta}$, large direct growth of $C_1$ dominates. 
(Unlike~\cite{Timme2010}, which requires overtaking until only two components remain, here the discontinuous transition occurs when there are still an infinite number of components in the limit of number of nodes $n \rightarrow \infty$.)

Using the notation of~\cite{BF}, let $t$ denote the number of accepted edges and $u$ the total number of sampled edges.  Thus for any $k$ (the maximum allowed component size), the fraction of accepted edges $t/u \ge g(k)$.  If $t/u$ is sufficiently large an edge leading to $C_1 n > k$ can be simply rejected.  In contrast, if $t/u=g(k)$ and the next edge sampled, denoted $e_{u+1}$, would lead to $C_1 n > k$ that edge cannot be simply rejected since $t/(u+1)<g(k)$.  One of two situations must happen, either: (i) $k$ increases until the edge $e_{u+1}$ is accommodated, or (ii) $g(k)$ decreases sufficiently that $t/(u+1) \ge g(k)$ and $e_{u+1}$ is rejected.  To determine which situation happens first, we need to determine the order of the smallest augmentation of $k$ that makes $t/(u+1)>g(k)$. 
%
For $\beta \ge 0.5$ the smallest fraction of accepted edges for any $k$ is
\begin{equation}
\frac{t}{u}= g(k) = \frac{1}{2}+\left(\frac{1}{2k}\right)^{\beta}.
\label{eqn:a}
\end{equation} 
Rearranging Eq.~\ref{eqn:a} and differentiating by $k$ yields 
\begin{equation}
\frac{du}{dk}=\frac{\beta}{2^{\beta}[1/2+{(1/2k)}^{\beta}]^{2}}\frac{t}{k^{\beta+1}}
\label{eqn:b}
\end{equation}
At some point in the subcritical evolution, $t \sim \mathcal{O}(n)$. (To build a component of size $\mathcal{O}(n)$ requires at least $\mathcal{O}(n)$ edges.) 
For the BFW model with $\beta=0.5$ it has been established rigorously that by the end of stage $k=25$, $t/n\rightarrow 0.841$ as $n\rightarrow\infty$~\cite{BF}. We establish via numerical simulation that $t/n > 0.82$ by the end stage $k=25$ for the full range $0.5 < \beta < 10$.  
Plugging  $t \sim \mathcal{O}(n)$ into Eq.~\ref{eqn:b}, once $k\sim n^{\gamma}$  (with $\gamma < 1$ for the subcritical region) then ${du}/{dk}\sim n^{1-\gamma\beta-\gamma}$. Thus an increase in $k$ of order $\mathcal{O}(n^{\gamma\beta+\gamma-1})$ is sufficient to ensure $t/(u+1)\geq g(k)$ and that edge $e_{u+1}$ can be rejected. 
But there are different behaviors for $\beta > 1$ and $\beta < 1$.

For $\beta>1$ there are three regimes.
$(i)$ For $\gamma \leq 1/(\beta+1)$ then $\gamma\beta+\gamma-1 \le 0$ so the necessary $\mathcal{O}(n^{\gamma\beta+\gamma-1})$ increase in $k$  requires only $k\rightarrow k+1$.   $(ii)$ Then once $1/(\beta+1) < \gamma < 1/\beta$,  an increase in $k$ of $\mathcal{O}(n^{\gamma\beta+\gamma-1}) < k \sim C_1 n$ is required.  $(iii)$ However, once $\gamma>1/\beta$, then $\mathcal{O}(n^{\gamma\beta+\gamma-1}) > C_1 n$.
Here the required increase in $k$ is greater than $C_1n$, allowing $C_1$ to even double in size, and edge $e_{u+1}$ must be accepted. 
So, 
once in the regime $C_1 n \sim n^{1/\beta}$ every time  $t/u=g(k)$ the next 
edge, $e_{u+1}$, must be accepted. 
In this situation, the probability two components are merged becomes, as in \ER~\cite{ER}, proportional to the product of their sizes. 

\begin{figure}
 
 \centering
\includegraphics[width=0.49\textwidth]{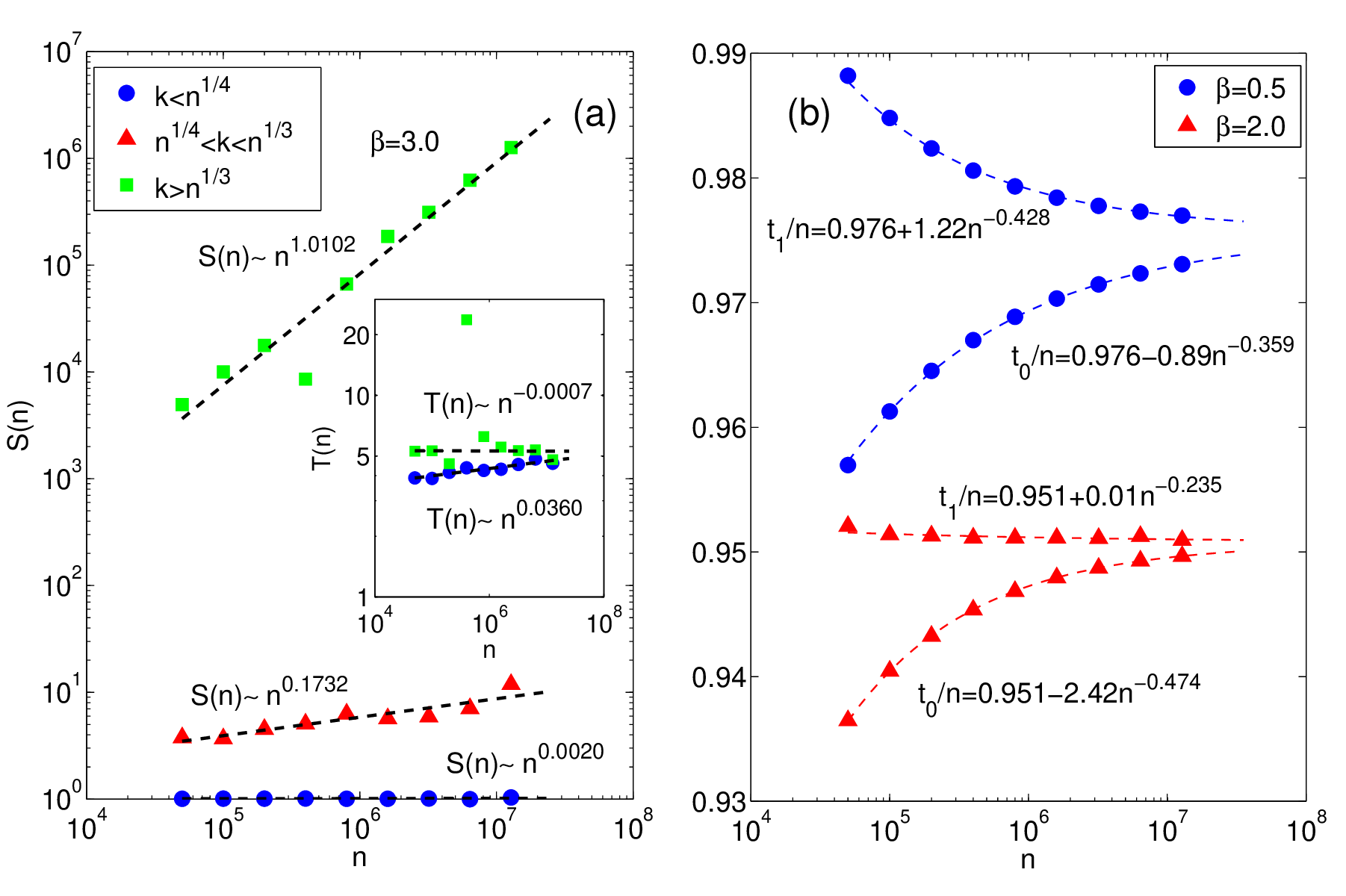}
 \caption{(a) The analogous plot to Fig.~\ref{fig:bs}(c), but with $\beta=3$: once $k>n^{1/\beta}$, the largest component merges with other macroscopic components ($S(n)\sim n^{1.0102}$).
 (b) Bounding the critical window from above and below to estimate $t_c$. For each $\beta$ value the lower line shows the largest value of $t$ for which $C_1 < n^{1/2}$, and the upper line the smallest value of $t$ for which $C_1 > 0.5 n$ for $\beta=0.5$ and $C_1>0.55n$ for $\beta=2$, yielding $t_c\approx 0.976n$ for $\beta=0.5$ and $t_c\approx 0.951$ for $\beta=2$. 
 }
 \label{fig:beta3tc}
 \end{figure}

For $\beta\in[0.5,1)$ there are only two regimes. 
$(i)$ Here again if $\gamma < 1/(\beta+1)$ then $k\rightarrow k+1$ allows edge $e_{u+1}$ to be rejected. 
$(ii)$ This regime extends until
$\gamma \geq 1/(\beta+1)$, when $\gamma\beta+\gamma-1 \ge 0$, but now we use the less strict property that 
$\gamma\beta+\gamma-1\leq \beta$ and thus $n^{\beta\gamma+\gamma-1}\le n^\beta < k$.
So throughout the evolution a sub-linear increase in $k$  of at most $n^\beta$ allows edge $e_{u+1}$ to be rejected. The slow increase results in multiple components of size similar to $C_1$ throughout the evolution. In particular, 
once $C_1n=\delta n$ with $\delta \ll 1$,  there exist many components of size $\mathcal{O}(n)$.  
Order them as $C_{1}n\geq C_{2}n...\geq C_{l}n$. 
Assuming $``>"$ strictly holds (\ie, choosing only one component of each size in the case of degenerate sized components), 
there will be components $C_l, C_{l-1}$ such that $C_{l}+C_{l-1}>C_{1}$.  (If instead $C_{l}+C_{l-1}\le C_{1}$, the two smaller components would merge together very quickly as the probability of randomly sampling an edge that connects them at any step $u$ is $C_l(u)C_{l-1}(u)$, of size $\mathcal{O}(1)$, and the edge would be accepted since $C_{l}+C_{l-1}\le C_1 \le k/n$.)
Due to the slow increase in $k$, which is in increments of at most $\mathcal{O}(n^\beta)$, there will be a point when $C_1n<(C_l+C_{l-1})n < k < (C_l + C_{l-2})n$, 
allowing for growth by overtaking, when $C_l$ and $C_{l-1}$ merge to become the new $C_1$, what was $C_1$ becomes $C_2$, and what was $C_{l-2}$ becomes $C_{l-1}$.  
%
The overtaking mechanism allows several large components to grow to the same order in size which is a necessary condition to generate a strongly discontinuous percolation transition. 
We explicitly observe this overtaking process for $\beta=0.5$ 
as shown inset in Fig.~\ref{fig:as}(a).

\begin{figure}
 
 \centering
\includegraphics[width=0.49\textwidth]{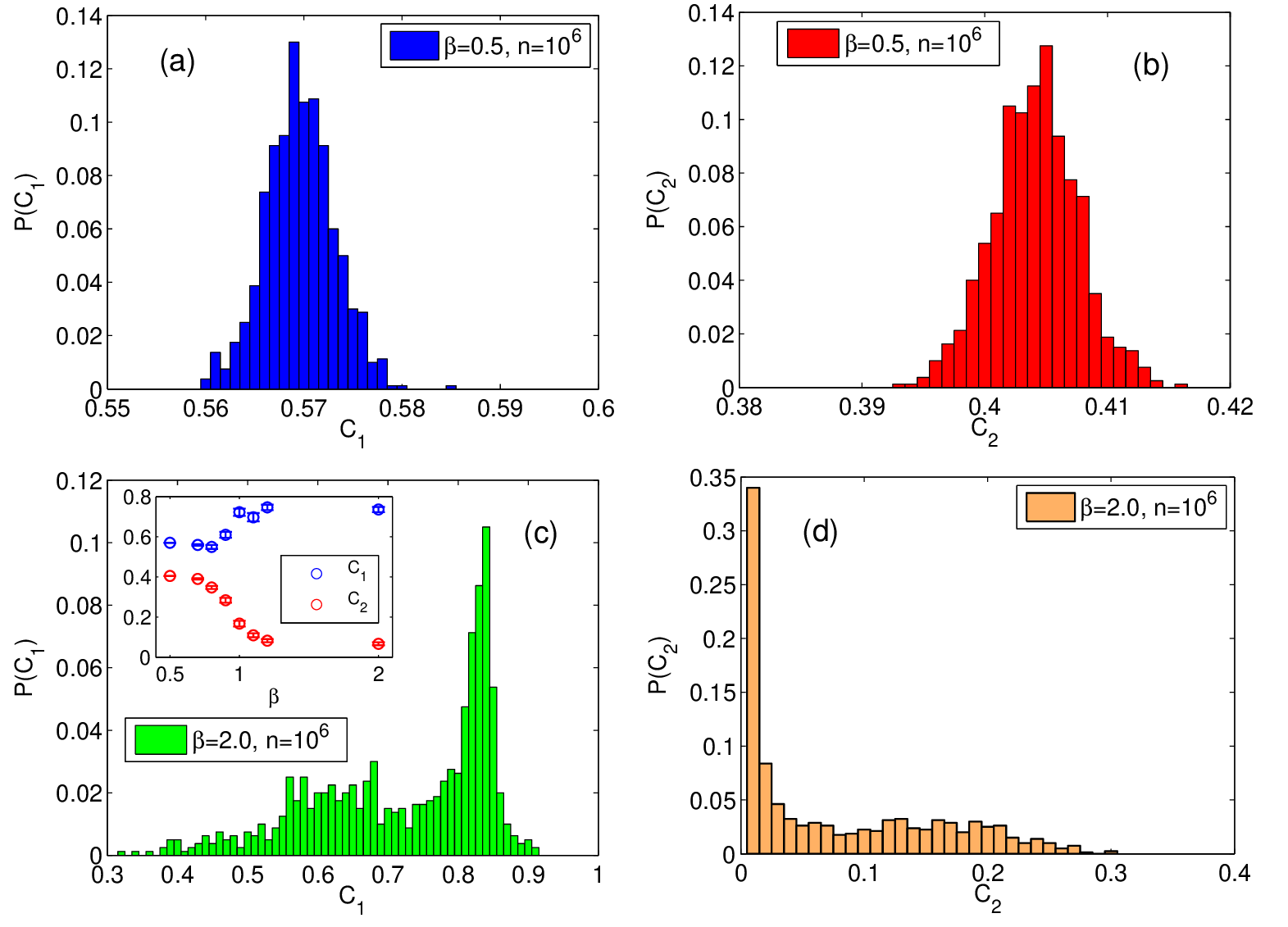}
 \caption{(a-d) Distribution in values of $C_1$ and $C_2$ at $t_c$ obtained over 100 independent realizations for $n=10^{6}$. ($t_c$ is defined as the single edge whose addition causes the biggest increase in $C_1$.) (a) $C_1$ for $\beta=0.5$. (b) $C_2$ for $\beta=0.5$. (c) $C_1$ for $\beta=2.0$. (d) $C_2$ for $\beta=2.0$.
Inset to (c) shows average size of $C_{1}$ and $C_{2}$ at $t_c$ over 100 realizations for different $\beta$. 
}
 \label{fig:cs}
 \end{figure}

We confirm these predictions via numerical simulations using two choices, $\beta=1/2$ and $\beta=2.0$.  
%
Let $S(n)$ denote the average {\it size} of the component $C_{i}$ which connects to $C_{1}$ via direct growth for a system of size $n$, and let $T(n)$ denote the number of {\it times} direct growth occurs. 
Figure~\ref{fig:bs}(b) is for $\beta=1/2$ where our analysis predicts two regimes separated by $k=n^{1/(\beta+1)}=n^{2/3}$. As shown in the inset, throughout both regimes $S(n)\approx 1.1$ is essentially a positive constant.  But $T(n)$ (the main figure) shows a distinct regime change. At first $T(n) \sim n^{0.19}$. Then in the  regime starting with $k=n^{2/3}$ up to and including  $t_c$ we see a much more rapid increase, $T(n) \sim n^{0.51}$.  So we see direct growth occurring more frequently in the second regime, but the direct growth continues to be due to merging with an essentially isolated node.

Figure~\ref{fig:bs}(c) is for $\beta=2$, where our analysis predicts three regimes. 
Up until $k= n^{1/(\beta+1)} = n^{1/3}$ the behavior is the same as for $\beta=0.5$ as expected since $\Delta k=1$ is enough for an edge to be rejected and we see $S(n) \approx 1.1$.  Then in the second regime of $n^{1/3} < k < n^{1/2}$, $S(n)$ is larger and has large fluctuations.  Finally in the third regime starting from $k=n^{1/\beta}=n^{1/2}$ up until edge $t_c$, we  see $C_1$ grow in large bursts, with $S(n)\sim n^{0.99}$, so $C_1$ merges with other essentially macroscopic  components.   As shown inset, $T(n)$ is essentially independent of regimes, with $T(n) \approx 5$ in the first regime and in the third, with $T(n) \le 1$  but fluctuating in the second (not shown).  The analogous three regimes and behaviors for $\beta=3$ are shown in Fig.~\ref{fig:beta3tc}(a). Here, once $k=n^{1/\beta}=n^{1/3}$, we see $C_1$ grow directly in large bursts with $S(n)\sim n^{1.01}$.


\begin{figure*}
 
\includegraphics[width=0.96\textwidth]{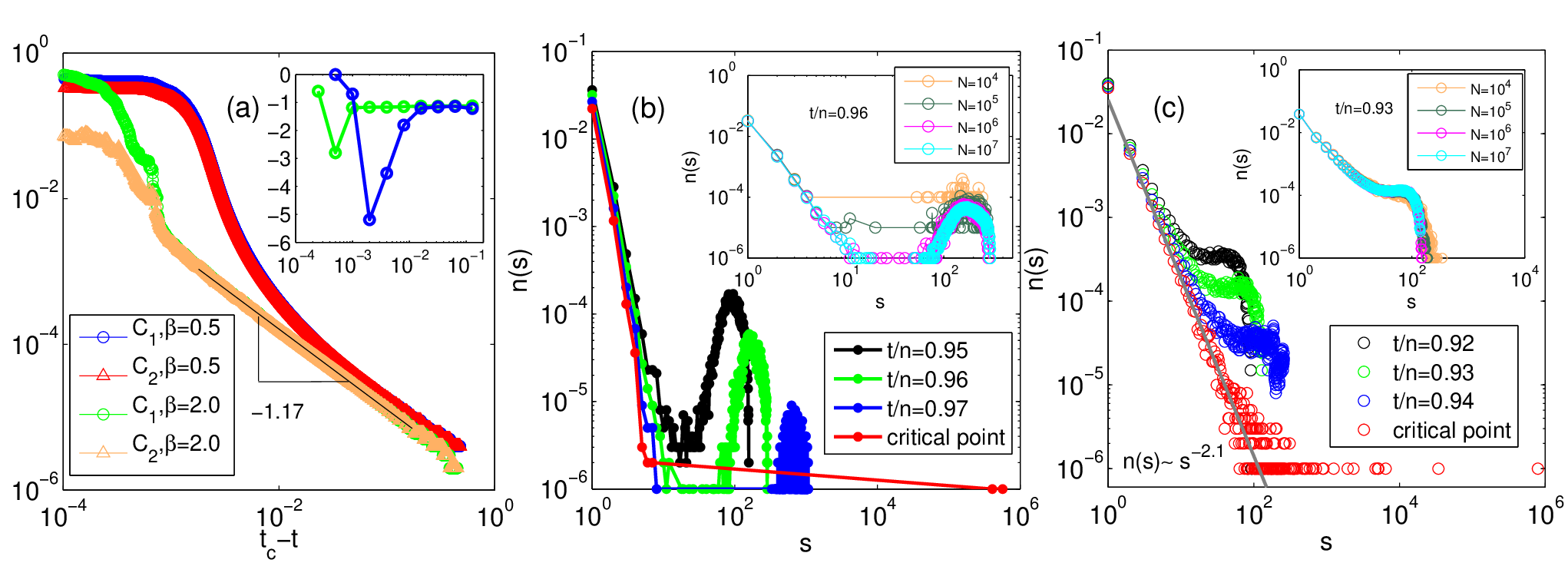}
\centering
\caption{No evident scaling behaviors for $\beta=0.5$ whereas quantities for $\beta=2.0$ exhibit critical scaling.   
(a)  $C_{1}$ and $C_{2}$ versus $t_{c}-t$.  For $\beta=2.0$ both $C_1,C_2 \sim (t_c-t)^{-1.17}$, yet $\beta=0.5$ shows no obvious scaling. Inset is the local slope estimate for $C_{1}$.
(b) Distribution of component density $n(s)$ (number of components of size $s$ divided by $n$) at different points in the evolution for $\beta=0.5$.   Inset is $n(s)$ at $t/n=0.96$ for various $n$, showing no finite size effects in the location of the right hump.  
(c) Evolution of $n(s)$ for $\beta=2.0$, with $n(s) \sim s^{-2.1}$ at $t_c$.  Inset is $n(s)$ at $t/n=0.93$ for various $n$, again showing no finite size effects.}
%
%
 \label{fig:ds}
 \end{figure*}

For $\beta=2.0$,  due to linear increase permitted in $k$ once $k>n^{1/2}$ and hence acceptance of random edges, we observe large fluctuations in the size of the giant components at $t_c$. For $\beta=0.5$,  due to the slow sub-linear increase in $k$,  
the sizes of the components evolve in a predictable manner.  Figure~\ref{fig:cs} shows these behaviors, with (a) showing $C_1$ and (b)  $C_2$ observed at $t_c$ over 100 independent realizations for $\beta=0.5$ and (c) and (d) the equivalent for $\beta=2$.  
Note for $\beta=0.5$, $t_c\approx 0.976n$ and for $\beta=2.0, t_c\approx 0.951n$, 
as shown  in Fig.~\ref{fig:beta3tc}(b). 


A scaling analysis of the general BFW model illuminates other unique features. With $\beta=2.0$, BFW exhibits critical scaling distinct from \ER~(ER)\cite{ER} and with no finite size effects. While for $\beta=0.5$ the model does not exhibit either critical scaling or finite size effects. We first examine $C_1$ and $C_2$ near $t_c$, as shown in Fig.~\ref{fig:ds}(a). For $\beta=2$ we find 
 $C_{1}, C_{2}\sim (t_{c}-t)^{-\eta}$ with $\eta=1.17$, the same scaling of $C_2$ as for the Product Rule (PR), 
a fixed choice 
rule studied in~\cite{EPScience},~\cite{RaissaMichael,RadicchiPRE2010}. 
We also consider standard finite size scaling $C_1 = n^{-\gamma/\nu} F \left[\left( t - t_c\right) n^{1/\nu} \right]$ and perform a data collapse to determine $1/\nu=0.49\pm0.02$ for BFW with $\beta=2$. Note, for ER,  $1/\nu = 1/3$.
For BFW with $\beta=0.5$, $C_{1},$ and $C_{2}$ show no obvious scaling behavior.

More importantly we study the component size density $n(s)$ (the number of components of size $s$ divided by $n$). We measure the distribution of $n(s)$ at different points in the evolution up to the critical point.   For $\beta=2.0$, Fig.~\ref{fig:ds}(c), the behavior is similar to that for PR and other edge competition models with fixed choice, where at the critical point there is clear scaling behavior, $n(s)\sim s^{-\tau}$ with $\tau=2.1$ (the same $\tau$ as for PR; $\tau=2.5$ for ER).  
Yet, as shown in Fig.~\ref{fig:ds}(b), the evolution for $\beta=0.5$ does not show any scaling. 
There is a pronounced right-hump which forms early in the evolution, then moves rightward due to overtaking until there are only two large components remaining at $t_c$.
Inset to Figs.~\ref{fig:ds}(b) and (c), respectively, are $n(s)$ at $t=0.96$ and $t=0.93$  for many different values of $n$. The peak of the right-hump is independent of $n$.  

\begin{figure}
 
 \centering
\includegraphics[width=0.45\textwidth]{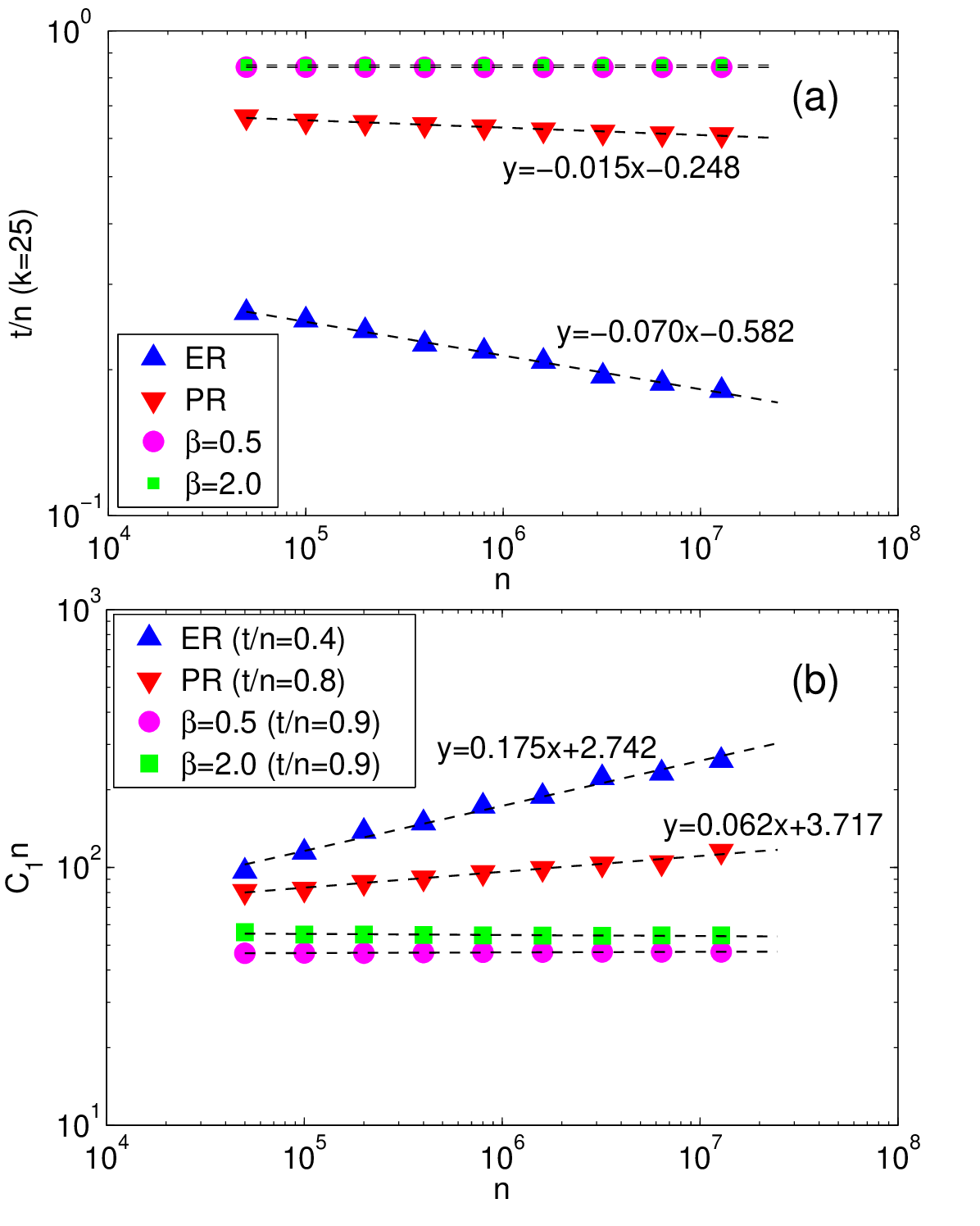}
 \caption{
 Lack of finite size effects for BFW with $\beta=0.5$ and $\beta=2.0$. (a) Fraction of added edges, $t/n$, once $C_1n=k=25$, versus  system size $n$, for ER, PR, and BFW with $\beta=0.5$ and $\beta=2.0$.  (b) Size of largest component ($C_1n$) versus  system size $n$ at the time when a specified fraction of edges have been added for ER, PR, and BFW with $\beta=0.5$ and $\beta=2.0$.
 }
 \label{fig:FS}
 \end{figure}

The BFW model with either $\beta=0.5$ or 2 shows no finite size effects, unlike ER and PR. First, for PR the location of the peak moves rightward with $n$, as  shown in~\cite{LeeKimPark} where a finite size scaling function for PR is established.   
Second, Fig.~\ref{fig:FS}(a) shows the fraction of edges added at the first time that $C_1n=25$ (denoted by $\frac{t}{n}(C_1n=25)$) versus $n$ for BFW, ER and  PR.  For BFW this value is independent of system size and converges to a positive constant for both $\beta=0.5$ and $2.0$, whereas it decreases to $0$ asymptotically for ER and PR, with $\frac{t}{n}(C_1n=25)\sim n^{-\tau}$, $\tau=0.070$ and $0.015$ respectively. Finally, rather than measuring $t/n$ for fixed $C_1n$, we can measure the value of $C_1n$ at the time when $t/n$ attains a specified value. Figure~\ref{fig:FS}(b) shows that for BFW with both $\beta=0.5$ and  $2.0$, $C_1 n $ is a positive constant for $t/n=0.9$.  Whereas for ER and PR then $C_{1}n\sim n^{\theta}$ with $\theta=0.175, 0.062$ respectively, measured in the subcritical regime for each model respectively ($t/n=0.4$ for ER, and $t/n=0.8$ for PR.)


\section{Summary}
In summary, we have derived the underlying mechanism that leads to the discontinuous percolation transition of the BFW model. 
This mechanism of growth by overtaking is a common mechanism observed in economic and ecological systems~\cite{overtakingBusiness,overtakingEco,overtakingEcon}.   
We have previously shown that by varying the asymptotic fraction of accepted edges, we can control the number of resulting giant components~\cite{CD2011}. In particular we studied a BFW model with an acceptance function $g(k) = \alpha + (2k)^{-1/2}$ (in the discontinuous regime since $\beta=1/2$) and showed that $\alpha$ controls the number of resulting giants. 
 These giant components are stable and persist throughout the supercritical evolution: Once in the supercritical regime, there are always sufficient edges internal to components sampled that whenever an edge connecting two giant components is sampled it can be rejected. The same simple analysis holds  for the most general BFW model, with $g(k) = \alpha + (2k)^{-\beta}$.  From a practical perspective, we now have an algorithm for generating a specified number of stable giant components in either a discontinuous or continuous percolation transition.  
From a theoretical perspective, we now have an analytic understanding of the growth mechanism underlying 
the BFW model, which leads to a discontinuous percolation transition and to multiple stable giant components. Note, growth by overtaking is one mechanism that gives rise to discontinuous percolation, there are also other mechanisms such as cooperation~\cite{Bhizani}.

The mathematical analysis herein strongly suggests the existence of a tricritical point at $\beta=1$, yet we cannot currently access this regime numerically.  Due to the finite  system size, for  $\beta\in[0.7, 1.0]$ we occasionally see significant direct growth of the largest component in simulations.  The rate of direct growth decreases with system size, but our current systems of size $10^7$ are to small to allow a quantitative study. This is an outstanding challenge.

\acknowledgments
This work was supported in part by the Defense Threat Reduction Agency, Basic Research Award No. HDTRA1-10-1-0088,  the Army Research Laboratory under Cooperative Agreement Number W911NF-09-2-0053, the National Basic Research Program of China (No. 2005CB321902) and the National Grand Fundamental Research 973 Program of China (No. 2013CB329602).

\end{document}